\documentclass[conference,10pt,twocolumn]{IEEEtran}

\usepackage{cite}
\usepackage{graphicx}
\usepackage{amssymb}
\usepackage{amsmath}
\usepackage{subfigure}
\usepackage{booktabs}
\usepackage{amsthm}
\usepackage{multirow}
\usepackage{microtype}
\usepackage{tablefootnote}
\usepackage{balance}
\usepackage{xcolor}
\usepackage{hyperref}

\usepackage{amssymb}
\usepackage{amsfonts}
\usepackage{mathrsfs}
\usepackage{xspace}
\usepackage{bm}
\usepackage{upgreek}

\newcommand{\safemath}[2]{\newcommand{#1}{\ensuremath{#2}\xspace}}

\safemath{\bma}{\mathbf{a}}
\safemath{\bmb}{\mathbf{b}}
\safemath{\bmc}{\mathbf{c}}
\safemath{\bmd}{\mathbf{d}}
\safemath{\bme}{\mathbf{e}}
\safemath{\bmf}{\mathbf{f}}
\safemath{\bmg}{\mathbf{g}}
\safemath{\bmh}{\mathbf{h}}
\safemath{\bmi}{\mathbf{i}}
\safemath{\bmj}{\mathbf{j}}
\safemath{\bmk}{\mathbf{k}}
\safemath{\bml}{\mathbf{l}}
\safemath{\bmm}{\mathbf{m}}
\safemath{\bmn}{\mathbf{n}}
\safemath{\bmo}{\mathbf{o}}
\safemath{\bmp}{\mathbf{p}}
\safemath{\bmq}{\mathbf{q}}
\safemath{\bmr}{\mathbf{r}}
\safemath{\bms}{\mathbf{s}}
\safemath{\bmt}{\mathbf{t}}
\safemath{\bmu}{\mathbf{u}}
\safemath{\bmv}{\mathbf{v}}
\safemath{\bmw}{\mathbf{w}}
\safemath{\bmx}{\mathbf{x}}
\safemath{\bmy}{\mathbf{y}}
\safemath{\bmz}{\mathbf{z}}
\safemath{\bmzero}{\mathbf{0}}
\safemath{\bmone}{\mathbf{1}}

\bmdefine{\biad}{a}
\bmdefine{\bibd}{b}
\bmdefine{\bicd}{c}
\bmdefine{\bidd}{d}
\bmdefine{\bied}{e}
\bmdefine{\bifd}{f}
\bmdefine{\bigd}{g}
\bmdefine{\bihd}{h}
\bmdefine{\biid}{i}
\bmdefine{\bijd}{j}
\bmdefine{\bikd}{k}
\bmdefine{\bild}{l}
\bmdefine{\bimd}{m}
\bmdefine{\bind}{n}
\bmdefine{\biod}{o}
\bmdefine{\bipd}{p}
\bmdefine{\biqd}{q}
\bmdefine{\bird}{r}
\bmdefine{\bisd}{s}
\bmdefine{\bitd}{t}
\bmdefine{\biud}{u}
\bmdefine{\bivd}{v}
\bmdefine{\biwd}{w}
\bmdefine{\bixd}{x}
\bmdefine{\biyd}{y}
\bmdefine{\bizd}{z}

\bmdefine{\bixid}{\xi}
\bmdefine{\bilambdad}{\lambda}
\bmdefine{\bimud}{\mu}
\bmdefine{\bithetad}{\theta}
\bmdefine{\biphid}{\phi}
\bmdefine{\bideltad}{\delta}

\safemath{\bmia}{\biad}
\safemath{\bmib}{\bibd}
\safemath{\bmic}{\bicd}
\safemath{\bmid}{\bidd}
\safemath{\bmie}{\bied}
\safemath{\bmif}{\bifd}
\safemath{\bmig}{\bigd}
\safemath{\bmih}{\bihd}
\safemath{\bmii}{\biid}
\safemath{\bmij}{\bijd}
\safemath{\bmik}{\bikd}
\safemath{\bmil}{\bild}
\safemath{\bmim}{\bimd}
\safemath{\bmin}{\bind}
\safemath{\bmio}{\biod}
\safemath{\bmip}{\bipd}
\safemath{\bmiq}{\biqd}
\safemath{\bmir}{\bird}
\safemath{\bmis}{\bisd}
\safemath{\bmit}{\bitd}
\safemath{\bmiu}{\biud}
\safemath{\bmiv}{\bivd}
\safemath{\bmiw}{\biwd}
\safemath{\bmix}{\bixd}
\safemath{\bmiy}{\biyd}
\safemath{\bmiz}{\bizd}

\safemath{\bmxi}{\bixid}
\safemath{\bmlambda}{\bilambdad}
\safemath{\bmmu}{\bimud}
\safemath{\bmtheta}{\bithetad}
\safemath{\bmphi}{\biphid}
\safemath{\bmdelta}{\bideltad}

\safemath{\bA}{\mathbf{A}}
\safemath{\bB}{\mathbf{B}}
\safemath{\bC}{\mathbf{C}}
\safemath{\bD}{\mathbf{D}}
\safemath{\bE}{\mathbf{E}}
\safemath{\bF}{\mathbf{F}}
\safemath{\bG}{\mathbf{G}}
\safemath{\bH}{\mathbf{H}}
\safemath{\bI}{\mathbf{I}}
\safemath{\bJ}{\mathbf{J}}
\safemath{\bK}{\mathbf{K}}
\safemath{\bL}{\mathbf{L}}
\safemath{\bM}{\mathbf{M}}
\safemath{\bN}{\mathbf{N}}
\safemath{\bO}{\mathbf{O}}
\safemath{\bP}{\mathbf{P}}
\safemath{\bQ}{\mathbf{Q}}
\safemath{\bR}{\mathbf{R}}
\safemath{\bS}{\mathbf{S}}
\safemath{\bT}{\mathbf{T}}
\safemath{\bU}{\mathbf{U}}
\safemath{\bV}{\mathbf{V}}
\safemath{\bW}{\mathbf{W}}
\safemath{\bX}{\mathbf{X}}
\safemath{\bY}{\mathbf{Y}}
\safemath{\bZ}{\mathbf{Z}}

\safemath{\bZero}{\mathbf{0}}
\safemath{\bOne}{\mathbf{1}}
\safemath{\bDelta}{\mathbf{\Delta}}
\safemath{\bLambda}{\mathbf{\UpLambda}}
\safemath{\bPhi}{\mathbf{\Upphi}}
\safemath{\bSigma}{\mathbf{\Upsigma}}
\safemath{\bOmega}{\mathbf{\Upomega}}
\safemath{\bTheta}{\mathbf{\Uptheta}}

\bmdefine{\biAd}{A}
\bmdefine{\biBd}{B}
\bmdefine{\biCd}{C}
\bmdefine{\biDd}{D}
\bmdefine{\biEd}{E}
\bmdefine{\biFd}{F}
\bmdefine{\biGd}{G}
\bmdefine{\biHd}{H}
\bmdefine{\biId}{I}
\bmdefine{\biJd}{J}
\bmdefine{\biKd}{K}
\bmdefine{\biLd}{L}
\bmdefine{\biMd}{M}
\bmdefine{\biOd}{N}
\bmdefine{\biPd}{O}
\bmdefine{\biQd}{P}
\bmdefine{\biRd}{R}
\bmdefine{\biSd}{S}
\bmdefine{\biTd}{T}
\bmdefine{\biUd}{U}
\bmdefine{\biVd}{V}
\bmdefine{\biWd}{W}
\bmdefine{\biXd}{X}
\bmdefine{\biYd}{Y}
\bmdefine{\biZd}{Z}

\bmdefine{\biDelta}{\Delta}
\bmdefine{\biLambda}{\Lambda}
\bmdefine{\biPhi}{\Phi}
\bmdefine{\biSigma}{\Sigma}
\bmdefine{\biOmega}{\Omega}
\bmdefine{\biTheta}{\Theta}

\safemath{\bimA}{\biAd}
\safemath{\bimB}{\biBd}
\safemath{\bimC}{\biCd}
\safemath{\bimD}{\biDd}
\safemath{\bimE}{\biEd}
\safemath{\bimF}{\biFd}
\safemath{\bimG}{\biGd}
\safemath{\bimH}{\biHd}
\safemath{\bimI}{\biId}
\safemath{\bimJ}{\biJd}
\safemath{\bimK}{\biKd}
\safemath{\bimL}{\biLd}
\safemath{\bimM}{\biMd}
\safemath{\bimN}{\biNd}
\safemath{\bimO}{\biOd}
\safemath{\bimP}{\biPd}
\safemath{\bimQ}{\biQd}
\safemath{\bimR}{\biRd}
\safemath{\bimS}{\biSd}
\safemath{\bimT}{\biTd}
\safemath{\bimU}{\biUd}
\safemath{\bimV}{\biVd}
\safemath{\bimW}{\biWd}
\safemath{\bimX}{\biXd}
\safemath{\bimY}{\biYd}
\safemath{\bimZ}{\biZd}

\safemath{\bimDelta}{\biDelta}
\safemath{\bimLambda}{\biLambda}
\safemath{\bimPhi}{\biPhi}
\safemath{\bimSigma}{\biSigma}
\safemath{\bimOmega}{\biOmega}
\safemath{\bimTheta}{\biTheta}

\safemath{\setA}{\mathcal{A}}
\safemath{\setB}{\mathcal{B}}
\safemath{\setC}{\mathcal{C}}
\safemath{\setD}{\mathcal{D}}
\safemath{\setE}{\mathcal{E}}
\safemath{\setF}{\mathcal{F}}
\safemath{\setG}{\mathcal{G}}
\safemath{\setH}{\mathcal{H}}
\safemath{\setI}{\mathcal{I}}
\safemath{\setJ}{\mathcal{J}}
\safemath{\setK}{\mathcal{K}}
\safemath{\setL}{\mathcal{L}}
\safemath{\setM}{\mathcal{M}}
\safemath{\setN}{\mathcal{N}}
\safemath{\setO}{\mathcal{O}}
\safemath{\setP}{\mathcal{P}}
\safemath{\setQ}{\mathcal{Q}}
\safemath{\setR}{\mathcal{R}}
\safemath{\setS}{\mathcal{S}}
\safemath{\setT}{\mathcal{T}}
\safemath{\setU}{\mathcal{U}}
\safemath{\setV}{\mathcal{V}}
\safemath{\setW}{\mathcal{W}}
\safemath{\setX}{\mathcal{X}}
\safemath{\setY}{\mathcal{Y}}
\safemath{\setZ}{\mathcal{Z}}
\safemath{\emptySet}{\varnothing}

\safemath{\colA}{\mathscr{A}}
\safemath{\colB}{\mathscr{B}}
\safemath{\colC}{\mathscr{C}}
\safemath{\colD}{\mathscr{D}}
\safemath{\colE}{\mathscr{E}}
\safemath{\colF}{\mathscr{F}}
\safemath{\colG}{\mathscr{G}}
\safemath{\colH}{\mathscr{H}}
\safemath{\colI}{\mathscr{I}}
\safemath{\colJ}{\mathscr{J}}
\safemath{\colK}{\mathscr{K}}
\safemath{\colL}{\mathscr{L}}
\safemath{\colM}{\mathscr{M}}
\safemath{\colN}{\mathscr{N}}
\safemath{\colO}{\mathscr{O}}
\safemath{\colP}{\mathscr{P}}
\safemath{\colQ}{\mathscr{Q}}
\safemath{\colR}{\mathscr{R}}
\safemath{\colS}{\mathscr{S}}
\safemath{\colT}{\mathscr{T}}
\safemath{\colU}{\mathscr{U}}
\safemath{\colV}{\mathscr{V}}
\safemath{\colW}{\mathscr{W}}
\safemath{\colX}{\mathscr{X}}
\safemath{\colY}{\mathscr{Y}}
\safemath{\colZ}{\mathscr{Z}}

\safemath{\opA}{\mathbb{A}}
\safemath{\opB}{\mathbb{B}}
\safemath{\opC}{\mathbb{C}}
\safemath{\opD}{\mathbb{D}}
\safemath{\opE}{\mathbb{E}}
\safemath{\opF}{\mathbb{F}}
\safemath{\opG}{\mathbb{G}}
\safemath{\opH}{\mathbb{H}}
\safemath{\opI}{\mathbb{I}}
\safemath{\opJ}{\mathbb{J}}
\safemath{\opK}{\mathbb{K}}
\safemath{\opL}{\mathbb{L}}
\safemath{\opM}{\mathbb{M}}
\safemath{\opN}{\mathbb{N}}
\safemath{\opO}{\mathbb{O}}
\safemath{\opP}{\mathbb{P}}
\safemath{\opQ}{\mathbb{Q}}
\safemath{\opR}{\mathbb{R}}
\safemath{\opS}{\mathbb{S}}
\safemath{\opT}{\mathbb{T}}
\safemath{\opU}{\mathbb{U}}
\safemath{\opV}{\mathbb{V}}
\safemath{\opW}{\mathbb{W}}
\safemath{\opX}{\mathbb{X}}
\safemath{\opY}{\mathbb{Y}}
\safemath{\opZ}{\mathbb{Z}}
\safemath{\opZero}{\mathbb{O}}
\safemath{\identityop}{\opI}

\safemath{\veca}{\bma}
\safemath{\vecb}{\bmb}
\safemath{\vecc}{\bmc}
\safemath{\vecd}{\bmd}
\safemath{\vece}{\bme}
\safemath{\vecf}{\bmf}
\safemath{\vecg}{\bmg}
\safemath{\vech}{\bmh}
\safemath{\veci}{\bmi}
\safemath{\vecj}{\bmj}
\safemath{\veck}{\bmk}
\safemath{\vecl}{\bml}
\safemath{\vecm}{\bmm}
\safemath{\vecn}{\bmn}
\safemath{\veco}{\bmo}
\safemath{\vecp}{\bmp}
\safemath{\vecq}{\bmq}
\safemath{\vecr}{\bmr}
\safemath{\vecs}{\bms}
\safemath{\vect}{\bmt}
\safemath{\vecu}{\bmu}
\safemath{\vecv}{\bmv}
\safemath{\vecw}{\bmw}
\safemath{\vecx}{\bmx}
\safemath{\vecy}{\bmy}
\safemath{\vecz}{\bmz}

\safemath{\veczero}{\bmzero}
\safemath{\vecone}{\bmone}
\safemath{\vecxi}{\bmxi}
\safemath{\veclambda}{\bmlambda}
\safemath{\vecmu}{\bmmu}
\safemath{\vectheta}{\bmtheta}
\safemath{\vecphi}{\bmphi}
\safemath{\vecdelta}{\bmdelta}

\safemath{\matA}{\bA}
\safemath{\matB}{\bB}
\safemath{\matC}{\bC}
\safemath{\matD}{\bD}
\safemath{\matE}{\bE}
\safemath{\matF}{\bF}
\safemath{\matG}{\bG}
\safemath{\matH}{\bH}
\safemath{\matI}{\bI}
\safemath{\matJ}{\bJ}
\safemath{\matK}{\bK}
\safemath{\matL}{\bL}
\safemath{\matM}{\bM}
\safemath{\matN}{\bN}
\safemath{\matO}{\bO}
\safemath{\matP}{\bP}
\safemath{\matQ}{\bQ}
\safemath{\matR}{\bR}
\safemath{\matS}{\bS}
\safemath{\matT}{\bT}
\safemath{\matU}{\bU}
\safemath{\matV}{\bV}
\safemath{\matW}{\bW}
\safemath{\matX}{\bX}
\safemath{\matY}{\bY}
\safemath{\matZ}{\bZ}
\safemath{\matzero}{\bmzero}

\safemath{\matDelta}{\bDelta}
\safemath{\matLambda}{\bLambda}
\safemath{\matPhi}{\bPhi}
\safemath{\matSigma}{\bSigma}
\safemath{\matOmega}{\bOmega}
\safemath{\matTheta}{\bTheta}

\safemath{\matidentity}{\matI}
\safemath{\matone}{\matO}

\safemath{\rnda}{A}
\safemath{\rndb}{B}
\safemath{\rndc}{C}
\safemath{\rndd}{D}
\safemath{\rnde}{E}
\safemath{\rndf}{F}
\safemath{\rndg}{G}
\safemath{\rndh}{H}
\safemath{\rndi}{I}
\safemath{\rndj}{J}
\safemath{\rndk}{K}
\safemath{\rndl}{L}
\safemath{\rndm}{M}
\safemath{\rndn}{N}
\safemath{\rndo}{O}
\safemath{\rndp}{P}
\safemath{\rndq}{Q}
\safemath{\rndr}{R}
\safemath{\rnds}{S}
\safemath{\rndt}{T}
\safemath{\rndu}{U}
\safemath{\rndv}{V}
\safemath{\rndw}{W}
\safemath{\rndx}{X}
\safemath{\rndy}{Y}
\safemath{\rndz}{Z}

\safemath{\rveca}{\bimA}
\safemath{\rvecb}{\bimB}
\safemath{\rvecc}{\bimC}
\safemath{\rvecd}{\bimD}
\safemath{\rvece}{\bimE}
\safemath{\rvecf}{\bimF}
\safemath{\rvecg}{\bimG}
\safemath{\rvech}{\bimH}
\safemath{\rveci}{\bimI}
\safemath{\rvecj}{\bimJ}
\safemath{\rveck}{\bimK}
\safemath{\rvecl}{\bimL}
\safemath{\rvecm}{\bimM}
\safemath{\rvecn}{\bimN}
\safemath{\rveco}{\bomO}
\safemath{\rvecp}{\bimP}
\safemath{\rvecq}{\bimQ}
\safemath{\rvecr}{\bimR}
\safemath{\rvecs}{\bimS}
\safemath{\rvect}{\bimT}
\safemath{\rvecu}{\bimU}
\safemath{\rvecv}{\bimV}
\safemath{\rvecw}{\bimW}
\safemath{\rvecx}{\bimX}
\safemath{\rvecy}{\bimY}
\safemath{\rvecz}{\bimZ}

\safemath{\rvecxi}{\bmxi}
\safemath{\rveclambda}{\bmlambda}
\safemath{\rvecmu}{\bmmu}
\safemath{\rvectheta}{\bmtheta}
\safemath{\rvecphi}{\bmphi}

\safemath{\rmatA}{\bimA}
\safemath{\rmatB}{\bimB}
\safemath{\rmatC}{\bimC}
\safemath{\rmatD}{\bimD}
\safemath{\rmatE}{\bimE}
\safemath{\rmatF}{\bimF}
\safemath{\rmatG}{\bimG}
\safemath{\rmatH}{\bimH}
\safemath{\rmatI}{\bimI}
\safemath{\rmatJ}{\bimJ}
\safemath{\rmatK}{\bimK}
\safemath{\rmatL}{\bimL}
\safemath{\rmatM}{\bimM}
\safemath{\rmatN}{\bimN}
\safemath{\rmatO}{\bimO}
\safemath{\rmatP}{\bimP}
\safemath{\rmatQ}{\bimQ}
\safemath{\rmatR}{\bimR}
\safemath{\rmatS}{\bimS}
\safemath{\rmatT}{\bimT}
\safemath{\rmatU}{\bimU}
\safemath{\rmatV}{\bimV}
\safemath{\rmatW}{\bimW}
\safemath{\rmatX}{\bimX}
\safemath{\rmatY}{\bimY}
\safemath{\rmatZ}{\bimZ}

\safemath{\rmatDelta}{\bimDelta}
\safemath{\rmatLambda}{\bimLambda}
\safemath{\rmatPhi}{\bimPhi}
\safemath{\rmatSigma}{\bimSigma}
\safemath{\rmatOmega}{\bimOmega}
\safemath{\rmatTheta}{\bimTheta}

\usepackage{amssymb}
\usepackage{amsfonts}
\usepackage{mathrsfs}
\usepackage{xspace}
\usepackage{bm}
\usepackage{fancyref}
\usepackage{textcomp}

\usepackage{multirow}
\usepackage{stmaryrd}

\newenvironment{textbmatrix}{	\setlength{\arraycolsep}{2.5pt}%
								\big[\begin{matrix}}{\end{matrix}\big]%
								\raisebox{0.08ex}{\vphantom{M}}}

\def\be{\begin{equation}}
\def\ee{\end{equation}}
\def\een{\nonumber \end{equation}}
\def\mat{\begin{bmatrix}}
\def\emat{\end{bmatrix}}
\def\btm{\begin{textbmatrix}}
\def\etm{\end{textbmatrix}}

\def\ba#1\ea{\begin{align}#1\end{align}}
\def\bas#1\eas{\begin{align*}#1\end{align*}}
\def\bs#1\es{\begin{split}#1\end{split}}
\def\bg#1\eg{\begin{gather}#1\end{gather}}
\def\bml#1\eml{\begin{multline}#1\end{multline}}
\def\bi#1\ei{\begin{itemize}#1\end{itemize}}

\newcommand{\lefto}{\mathopen{}\left}

\DeclareMathOperator{\Exop}{\opE}			

\newcommand{\Ex}[2]{\ensuremath{\Exop_{#1}\lefto[#2\right]}} 	

\newcommand{\tp}[1]{\ensuremath{#1^{T}}} 		
\newcommand{\herm}[1]{\ensuremath{#1^{H}}} 	
\newcommand{\inv}[1]{\ensuremath{#1^{-1}}} 	

\safemath{\dirac}{\delta}					
\safemath{\krond}{\dirac}					

\safemath{\upto}{\uparrow}
\safemath{\downto}{\downarrow}
\safemath{\iu}{j}							
\safemath{\ev}{\lambda}						
\safemath{\hilseqspace}{l^{2}}				
\newcommand{\banachfunspace}[1]{\setL^{#1}}	
\safemath{\hilfunspace}{\banachfunspace{2}}	
\newcommand{\floor}[1]{\lfloor #1 \rfloor}

\safemath{\SNR}{\textit{SNR}} 				
\safemath{\PAR}{\textit{PAR}} 				
\safemath{\No}{N_0}							
\safemath{\Es}{E_s}							
\safemath{\Eb}{E_b}							
\safemath{\EbNo}{\frac{\Eb}{\No}}
\safemath{\EsNo}{\frac{\Es}{\No}}

\DeclareMathOperator{\CHop}{\ensuremath{\opH}} 
\safemath{\tvir}{\rndh_{\CHop}}				
\safemath{\tvtf}{\rndl_{\CHop}}				
\safemath{\spf}{\rnds_{\CHop}}				
\safemath{\bff}{H_{\CHop}}					

\safemath{\ircf}{r_{h}}						
\safemath{\tftvcf}{r_{s}}					
\safemath{\tfcf}{r_{l}}						
\safemath{\bfcf}{r_{H}}						

\safemath{\tcorr}{c_h}						
\safemath{\scf}{c_{s}}						
\safemath{\tfcorr}{c_{l}}					
\safemath{\fcorr}{c_{H}}						

\safemath{\mi}{I}							
\safemath{\capacity}{C}						

\safemath{\normal}{\mathcal{N}}			
\safemath{\jpg}{\mathcal{CN}}			
\safemath{\mchain}{\leftrightarrow}		

\safemath{\dB}{\,\mathrm{dB}}
\safemath{\dBm}{\,\mathrm{dBm}}
\safemath{\Hz}{\,\mathrm{Hz}}
\safemath{\kHz}{\,\mathrm{kHz}}
\safemath{\MHz}{\,\mathrm{MHz}}
\safemath{\GHz}{\,\mathrm{GHz}}
\safemath{\s}{\,\mathrm{s}}
\safemath{\ms}{\,\mathrm{ms}}
\safemath{\mus}{\,\mathrm{\text{\textmu}s}}
\safemath{\ns}{\,\mathrm{ns}}
\safemath{\ps}{\,\mathrm{ps}}
\safemath{\meter}{\,\mathrm{m}}
\safemath{\mm}{\,\mathrm{mm}}
\safemath{\cm}{\,\mathrm{cm}}
\safemath{\m}{\,\mathrm{m}}
\safemath{\W}{\,\mathrm{W}}
\safemath{\mW}{\, \mathrm{mW}}
\safemath{\J}{\,\mathrm{J}}
\safemath{\K}{\,\mathrm{K}}
\safemath{\bit}{\,\mathrm{bit}}
\safemath{\nat}{\,\mathrm{nat}}

\safemath{\define}{\triangleq}			

\safemath{\equivalent}{\sim}
\safemath{\distas}{\sim}					
\safemath{\sdiff}{\Delta}				

\safemath{\reals}{\mathbb{R}}
\safemath{\positivereals}{\reals_{+}}
\safemath{\integers}{\mathbb{Z}}
\safemath{\posint}{\integers_{+}}
\safemath{\naturals}{\mathbb{N}}
\safemath{\posnaturals}{\naturals_{+}}
\safemath{\complexset}{\mathbb{C}}
\safemath{\rationals}{\mathbb{Q}}

\newcommand*{\fancyrefapplabelprefix}{app}		
\newcommand*{\fancyrefthmlabelprefix}{thm}		
\newcommand*{\fancyreflemlabelprefix}{lem}		
\newcommand*{\fancyrefcorlabelprefix}{cor}		
\newcommand*{\fancyrefdeflabelprefix}{def}		
\newcommand*{\fancyrefproplabelprefix}{prop}		
\newcommand*{\fancyrefexmpllabelprefix}{exmpl}
\newcommand*{\fancyrefalglabelprefix}{alg}		
\newcommand*{\fancyreftbllabelprefix}{tbl}		

\frefformat{vario}{\fancyrefseclabelprefix}{Section~#1}
\frefformat{vario}{\fancyrefthmlabelprefix}{Theorem~#1}
\frefformat{vario}{\fancyreftbllabelprefix}{Table~#1}
\frefformat{vario}{\fancyreflemlabelprefix}{Lemma~#1}
\frefformat{vario}{\fancyrefcorlabelprefix}{Corollary~#1}
\frefformat{vario}{\fancyrefdeflabelprefix}{Definition~#1}
\frefformat{vario}{\fancyreffiglabelprefix}{Figure~#1}
\frefformat{vario}{\fancyrefapplabelprefix}{Appendix~#1}
\frefformat{vario}{\fancyrefeqlabelprefix}{(#1)}
\frefformat{vario}{\fancyrefproplabelprefix}{Proposition~#1}
\frefformat{vario}{\fancyrefexmpllabelprefix}{Example~#1}
\frefformat{vario}{\fancyrefalglabelprefix}{Algorithm~#1}

\safemath{\dictab}{[\,\dicta\,\,\dictb\,]}

\safemath{\ysig}{\bmy}
\safemath{\ysighat}{\hat{\ysig}}
\safemath{\ysigdim}{M}
\safemath{\xsig}{\bmx}
\safemath{\xsigdim}{N}
\safemath{\nx}{n_x}
\safemath{\zsig}{\bmz}
\safemath{\zsigdim}{\ysigdim}
\safemath{\rsig}{\bmr}
\safemath{\Adict}{\bA}
\safemath{\Adicttilde}{\widetilde{\Adict}}
\safemath{\Adictdim}{\outputdim\times\xsigdim}
\safemath{\avec}{\bma}
\safemath{\avectilde}{\tilde{\avec}}
\safemath{\Bdict}{\bB}
\safemath{\Bdicttilde}{\widetilde{\Bdict}}
\safemath{\Cdict}{\bC}
\safemath{\cvec}{\bmc}
\safemath{\Ddict}{\bD}
\safemath{\Ddictdim}{\ysigdim\times\xsigdim}
\safemath{\dvec}{\bmd}
\safemath{\Ddicttilde}{\widetilde{\bD}}
\safemath{\Bonb}{\bB}
\safemath{\bvec}{\bmb}
\safemath{\Bonbdim}{\ysigdim\times\ysigdim}
\safemath{\noise}{\bmn}
\safemath{\noisedim}{\ysigim}
\safemath{\err}{\bme}
\safemath{\errdim}{\ysigdim}
\safemath{\errset}{\setE}
\safemath{\nerr}{n_e}
\safemath{\delop}{\bP_\errset}
\safemath{\delopc}{\bP_{{\errset}^c}}

\safemath{\cplxi}{\imath}
\safemath{\cplxj}{\jmath}

\safemath{\dict}{\matD}
\safemath{\inputdim}{N}		
\safemath{\outputdim}{M}		
\safemath{\sparsity}{S}	
\safemath{\inputdimA}{{N_a}}	
\safemath{\inputdimB}{{N_b}}	
\safemath{\elemA}{{n_a}}	
\safemath{\elemB}{{n_b}}	
\safemath{\resA}{\matR_a}	
\safemath{\resB}{\matR_b}	
\safemath{\subD}{\matS} 
\safemath{\subA}{\matS_a} 
\safemath{\subB}{\matS_b} 
\safemath{\dicta}{\matA} 	
\safemath{\dictb}{\matB} 	
\safemath{\hollowS}{H}
\safemath{\hollowA}{H_a}
\safemath{\hollowB}{H_b}
\safemath{\cross}{Z}
\safemath{\coh}{\mu_d}			
\safemath{\coha}{\mu_a}			
\safemath{\cohb}{\mu_b}			
\safemath{\mubs}{\nu}	
\safemath{\cohm}{\mu_m} 
\safemath{\dictset}{\setD}	
\safemath{\dictsetp}{\dictset(\coh,\coha,\cohb)}	
\safemath{\dictsetgen}{\dictset_\text{gen}}
\safemath{\dictsetgenp}{\dictsetgen(\coh)}
\safemath{\dictsetonb}{\dictset_\text{onb}}
\safemath{\dictsetonbp}{\dictsetonb(\coh)}

\safemath{\leftside}{U}
\safemath{\rightsideA}{R_a}
\safemath{\rightsideB}{R_b}

\safemath{\indexS}{\setI_S} 

\safemath{\na}{n_a}			
\safemath{\nb}{n_b}			
\safemath{\coeffa}{p_i}	
\safemath{\coeffb}{q_j}	
\safemath{\seta}{\setP}		
\safemath{\setb}{\setQ}     
\safemath{\setw}{\setW}	
\safemath{\setz}{\setZ}	
\safemath{\cola}{\veca}		
\safemath{\colb}{\vecb}		
\safemath{\cold}{\vecd}		
\safemath{\inputvec}{\vecx} 	
\safemath{\error}{\vece}	
\safemath{\noiseout}{\vecz} 	
\safemath{\inputvecel}{x}
\safemath{\inputveca}{\vecx_a}
\safemath{\inputvecb}{\vecx_b}
\safemath{\outputvec}{\vecy}	
\safemath{\lambdamin}{\lambda_{\mathrm{min}}}

\safemath{\elltwo}{\ell_2}
\safemath{\ellone}{\ell_1}
\safemath{\ellzero}{\ell_0}
\safemath{\ellinf}{\ell_\infty}
\safemath{\ellinftilde}{\ell_{\widetilde\infty}}
\safemath{\licard}{Z(\coh,\coha,\cohb)}
\safemath{\xsol}{\hat{x}}
\safemath{\xbord}{x_b}		
\safemath{\xstat}{x_s}		
\safemath{\xstatLone}{\tilde{x}_s}
\safemath{\order}{\mathcal{O}} 
\safemath{\scales}{\Theta} 
\safemath{\ones}{\mathbf{1}} 
\safemath{\zeroes}{\mathbf{0}} 
\safemath{\thlone}{\kappa(\coh,\cohb)} 
\safemath{\constoneA}{\delta} 
\safemath{\constoneB}{\epsilon} 
\safemath{\nlarge}{L}				   
\safemath{\sumlarge}{S_\nlarge}
\safemath{\maxlarger}{P_\nlarge}	   
\safemath{\Pzero}{\textrm{P0}}	
\safemath{\Pone}{\textrm{P1}}
\safemath{\vecfir}{\vecw}			 
\safemath{\vecsec}{\vecz}
\safemath{\elvecfir}{w}              
\safemath{\elvecsec}{z}				 
\safemath{\nlargefir}{n}
\safemath{\normout}{\gamma}
\safemath{\auxfun}{h}
\safemath{\supp}{\textrm{supp}}

\safemath{\indexa}{\ell}
\safemath{\indexb}{r}
\safemath{\indexc}{i}
\safemath{\indexd}{j}

\safemath{\project}{P}

\IEEEoverridecommandlockouts

\allowdisplaybreaks

\safemath{\Hj}{\bmj}
\safemath{\sj}{w}
\safemath{\Ej}{E_w}
\safemath{\quant}{Q}
\safemath{\compquant}{\mathcal{Q}}
\safemath{\Hest}{\hat{\bH}_{\text{est}}}
\safemath{\EIf}{\boldsymbol{\Lambda}}
\safemath{\rmmse}{\textit{RMSSE}}

\begin{document}
	
\title{Beam-Slicing for Jammer Mitigation \\ in mmWave Massive MU-MIMO}%
\author{\IEEEauthorblockN{Oscar Casta\~neda$^\ast$, Gian Marti$^\ast$, and Christoph Studer} \\[-0.0cm]
\IEEEauthorblockA{\em Department of Information Technology
and Electrical Engineering, ETH Z\"urich, Switzerland}
\thanks{$^\ast$OC and GM contributed equally to this work.}
\thanks{The work of OC and CS was supported in part by  ComSenTer, one of six centers in JUMP, a SRC program sponsored by DARPA. The work of CS was also supported by an ETH Research Grant and by the US National Science Foundation (NSF) under grants CNS-1717559 and ECCS-1824379.}\thanks{The authors thank D. Nonaca for helpful comments on this manuscript.}\thanks{Contact authors: OC and GM; e-mail: caoscar@ethz.ch, gimarti@ethz.ch}\\[-0.4cm]
}

\maketitle


\begin{abstract}
Millimeter-wave (mmWave) massive multi-user multiple-input multiple-output (MU-MIMO) technology promises unprecedentedly high data rates for next-generation wireless systems.
To be practically viable, mmWave massive MU-MIMO basestations (BS) must (i) rely on low-resolution data-conversion and (ii) be robust to jammer interference.
This paper considers the problem of mitigating the impact of a permanently transmitting jammer during uplink transmission to a BS equipped with low-resolution analog-to-digital converters (ADCs). To this end, we propose SNIPS, short for Soft-Nulling of Interferers with Partitions in Space.
SNIPS combines beam-slicing---a localized, analog spatial transform that focuses the jammer energy onto a subset of all ADCs---together with a soft-nulling data detector that exploits knowledge of which ADCs are contaminated by jammer interference.
Our numerical results show that SNIPS is able to successfully serve 65\% of the user equipments (UEs) for scenarios in which a conventional antenna-domain soft-nulling data detector is only able to serve 2\% of the UEs.
\end{abstract}


\section{Introduction} 
\label{sec:intro}

In order to meet the ever-growing demand for higher data rates, next-generation wireless communication systems are expected to rely on the vast amount of unused bandwidth available at millimeter wave (mmWave) frequencies~\cite{rappaport15a}.
Communication at mmWave frequencies is characterized by a high path loss that can be compensated for with massive multiple-input multiple-output (MIMO) technology.
Besides providing the basestation (BS) with a high array gain, massive MIMO also enables multi-user (MU) communication~\cite{larsson14a}.

The deployment of a BS equipped with a large number of antennas and corresponding radio-frequency (RF) chains poses implementation challenges in terms of system costs, power consumption, and circuit complexity.
A potential solution is to use low-resolution data converters that (i) reduce the power consumption of data conversion and (ii) simplify the linearity and noise requirements of the RF chain, which in turn translates into power consumption and circuit complexity savings~\cite{castaneda17a}.

Unfortunately, the use of low-resolution analog-to-digital converters (ADCs) leaves the BS vulnerable to jammers that could be introduced, for example, by a rogue user equipment (UE) or a malicious transmitter.
Previous works~\cite{yan14a,shen14a,kapetanovic13a,akhlaghpasand18a,vinogradova16a,zhao17a,do18a,akhlaghpasand20a,akhlaghpasand20b,bagherinejad21a} have analyzed the impact of different types of jamming attacks on massive MU-MIMO systems and proposed mitigation methods based on digital equalization.
However, none of these works have taken into consideration the compounding challenge of low-resolution data conversion: 
A jammer can either saturate low-resolution ADCs or (if gain-control is used) widen their quantization range to an extent that drowns  the signal in quantization noise, and thereby introduce distortions that are difficult to remove with subsequent digital equalization. 

\subsection{Contributions}
In this work, we propose novel means to limit the impact of a single, permanently transmitting jammer on mmWave massive MU-MIMO systems while taking into account the nonlinear distortions caused by ADCs.
Our numerical results indicate that, as long as the ADCs have a resolution that is sufficiently high (e.g., 8 bits for a 25\,dB jammer),
the jammer can be effectively mitigated by using a linear equalizer in the digital domain that utilizes an estimate of the jammer covariance matrix.
However, practical deployments of massive MU-MIMO are likely to rely on \mbox{low-resolution} ADCs.
In such situations, the jammer will force the ADCs' quantization range to drown the UE signals in quantization noise.
In order to enable jammer-robust communication with low-resolution ADCs, we propose a novel method called \textit{Soft-Nulling of Interferers with Partitions in Space (SNIPS)}.
SNIPS partitions the space using a non-adaptive, localized spatial transform prior to the data-conversion step.
We refer to this transform as \textit{beam-slicing}.
Thanks to the strong directionality of mmWave signals, beam-slicing focuses the jammer's energy onto a subset of all ADCs.
An estimate of how strong each ADC is affected by the jammer is then used by the data detector to generate improved estimates of the transmitted data. 

\subsection{Related Prior Work}

Several works have studied means that improve the resiliency of MIMO systems against jamming attacks.
These works have considered different attacks, such as constant jamming attacks~\cite{yan14a,shen14a}, in which the jammer is permanently transmitting, as well as other types of attacks in which the jammer transmits only at specific time instances, such as when the UEs transmit~\cite{yan14a} or during pilot transmission~\cite{kapetanovic13a}.
Furthermore, due to the complexity of the jammer problem, some works have devoted themselves only to detecting the presence of a jammer~\cite{kapetanovic13a,akhlaghpasand18a}, while other works have proposed methods to suppress jamming~\cite{yan14a,shen14a,vinogradova16a,zhao17a,do18a,akhlaghpasand20a,akhlaghpasand20b,bagherinejad21a}.

In our work, we focus on mitigating the interference of a permanently transmitting jammer. We now describe existing approaches that are related to handling the jammer's interference.
In~\cite{yan14a}, a method which uses the angle of arrival of the interfering signal for projecting the receive vector onto the subspace orthogonal to the interference is proposed for small-scale MIMO systems.
Also in the context of small-scale MIMO, reference~\cite{shen14a} proposes a method that uses differential encoding and exploits the ratio between channel coefficients.
Turning to massive MIMO systems, reference~\cite{vinogradova16a} uses random matrix theory to estimate the UEs' eigensubspace to then project the received signals onto that subspace, while~\cite{zhao17a} proposes methods which require perfect channel state information and cooperation between UEs and BS. 
Moreover, references~\cite{do18a,akhlaghpasand20a} utilize an estimate of the jammer channel to implement different versions of a jammer-robust zero-forcing detector.

Similarly to our work, references~\cite{akhlaghpasand20b,bagherinejad21a} propose to exploit spatially correlated channels in order to suppress jammer interference.
In particular, the work in~\cite{bagherinejad21a} applies a beamspace transform to separate the jammer from the UEs in the angular domain.
The beamspace transform \cite{brady13} is related to the concept of beam-slicing used by SNIPS.
As a matter of fact, the conventional beamspace transform (and even the absence of any spatial transform, i.e., the antenna domain) can be formulated as a special case of beam-slicing, and one can think in general of beam-slicing as a beamspace transform with a coarser angular resolution.
The key advantage of beam-slicing over the beamspace transform is that the former is composed of \textit{localized} transforms which only take inputs from a few adjacent antennas, making such transforms more amenable for analog circuit implementation~\cite{JoshiThesis}.
Another key difference of SNIPS is that it takes into account the effects of low-resolution quantization.
Most of the existing results do not consider the effects of hardware impairments---with one exception: 
Reference~\cite{akhlaghpasand20b} models the effects of hardware impairments (including quantization errors) as additive Gaussian noise, hence failing to accurately model the signal- and jammer-dependent distortions introduced by coarse ADCs.

\subsection{Notation}
Matrices and column vectors are represented by boldface uppercase and lowercase letters, respectively.
For a matrix $\bA$, the conjugate transpose is $\bA^H$, the $k$th column is $\bma_k$, and the Frobenius norm is $\| \bA \|_F$.
The $N\times N$ identity matrix and discrete Fourier transform (DFT) are $\bI_N$ and $\bF_N$, respectively, where $\herm{\bF_N}\bF_N=\bI_N$.
For a vector $\bma$, the $k$th entry is $a_k$, the $\ell_2$-norm is $\|\bma\|_2$, the real part is $\Re\{\bma\}$, and the imaginary part is $\Im\{\bma\}$.
Furthermore, $\text{diag}(\bma)$ is a diagonal matrix whose diagonal is formed by $\bma$.
Expectation with respect to the random vector~$\bmx$ is denoted by \Ex{\bmx}{\cdot}.
The floor function $\floor{x}$ returns the greatest integer less than or equal to $x$.
We define $i^2=-1$.


\section{Propagation Model}\label{sec:system}

We consider the uplink of a mmWave massive MU-MIMO system in which $U$ single-antenna UEs transmit data to a $B$ antenna BS, while a permanently transmitting, single-antenna jammer interferes with the BS receive signal. 
For this scenario, we consider the following frequency-flat input-output relation:
 \begin{equation}
\bmy = \bH\bms + \Hj\sj + \bmn. \label{eq:ant_io}
 \end{equation}
Here, $\bmy\in\complexset^B$ is the (unquantized) vector received by the BS antennas, $\bH\in\complexset^{B\times U}$ models the MIMO uplink channel matrix, $\bms\in\setS^U$ is the transmit vector whose (independent) entries correspond to the per-UE transmit symbols which take value in a constellation set $\setS$ (e.g., $16$-QAM), $\Hj\in\complexset^B$ is the channel from the jammer to the BS, $\sj\in\complexset$ is the jamming signal, and $\bmn\in\complexset^B$ is i.i.d.\ circularly-symmetric complex Gaussian noise with a per-entry variance of $\No$.
In what follows, we assume that the UE transmit symbols $s_u$, $u=1,\dots,U$, are independent and zero mean with variance $\Es$ so that $\Ex{\bms}{\bms\herm{\bms}}=\Es\bI_U$.
We model the jamming signal $\sj$ as circularly-symmetric complex Gaussian with variance $\Ej$.
All probabilistic quantities are assumed to be mutually independent. 


\section{SNIPS: Soft-Nulling of Interferers\\with Partitions in Space}
\label{sec:slice}

\begin{figure}[tp]
\centering
\includegraphics[width=0.9\columnwidth]{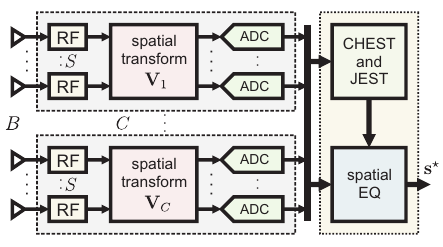}
\caption{System overview of SNIPS: The $B$ RF chains are clustered into~$C$ clusters of size $S$. The analog baseband signals are transformed cluster-wise to the beam-slice domain before being converted to the digital domain. SNIPS then performs jammer interference estimation, channel estimation, and data detection in digital beam-slice domain.}
\label{fig:system_overview}
\end{figure}

Our proposed method aims to protect most of the ADCs from the jammer by exploiting the strong spatial directivity of mmWave signals. 
Prior to analog-to-digital (A/D)-conversion, we apply a spatial transform that resolves the incident waves so that only a few ADCs are strongly affected by the jammer.
One may then discount the outputs of these jammer-distorted ADCs during equalization and detect the data symbols mainly based on the outputs of the distortion-free ADCs.

A na\"ive approach would be to transform the (unquantized) receive vector~$\bmy$ into the beamspace (or angular) domain~\cite{brady13} using a DFT according to $\bmy_{B}=\bF_B \bmy$, and set the entries of the beamspace vector $\bmy_{B}$ dominated by jammer interference to zero. 
Accordingly, the resulting vector $\bmy_{B,\text{mask}}$ will have entries equal to zero if they belong to jammer-contaminated beams, and otherwise equal to the corresponding entry in $\bmy_{B}$.
One could then equalize $\bmy_{B,\text{mask}}$ as if neither interference nor interference-cancellation had occurred, for instance with linear minimum mean-square error (LMMSE) estimation, 
\mbox{$\bms^\star = \inv{(\herm{\bH}\bH + \No/\Es\bI_B)}\herm{\bH}\bmy_{B\text{,mask}}$.}
While such an approach would be effective in suppressing jammers, implementing large spatial transforms in the analog domain is nontrivial, especially when considering hundreds of BS antennas~\cite{JoshiThesis}.

SNIPS is a jammer-mitigation method that relies on a distributed, localized, and hence small analog transform.
SNIPS does not discard the outputs of jammer-affected ADCs completely, but instead takes into account each ADC-output's fidelity by estimating the amount of jammer interference at the individual ADCs.
Moreover, SNIPS utilizes Bussgang's decomposition in order to take into account the effects of practical, low-resolution ADCs.
\fref{fig:system_overview} illustrates the complete SNIPS pipeline, which we detail hereafter.
 
\begin{figure}[tp]
\centering
\includegraphics[width=0.9\columnwidth]{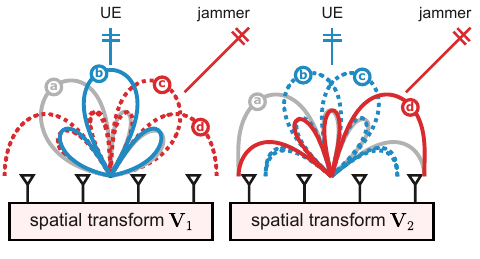}
\caption{Effect of $S\!=\!4$ beam-slicing for a $B\!=\!8$ antenna array considering one UE and one jammer (both in far-field). The spatial transform $\bV_1$ illustrates the $4$ beams (a), (b), (c), and (d) of $\bF_4$, of which (b) is perfectly aligned to capture the UE's transmitted power, while beams (c) and (d) partially capture the jammer's power. Such partial capturing would lead to unsatisfactory jammer mitigation. To increase angular diversity, the spatial transform $\bV_2$ first rotates the received signal so that the $4$ beams (a), (b), (c), and (d)  of the resulting transform are also shifted. As a result, the shifted beam (d) is now able to capture the jammer's power, which allows for improved jammer mitigation.}
\label{fig:explain_beamslicing}
\end{figure}

\subsection{Beam-Slicing} \label{subsec:beam_slicing}
Beam-slicing transforms partitions of the (unquantized) receive vector~$\bmy$ into shifted angular domains.
Beam-slicing is fully analog, non-adaptive, and operates in decentralized fashion. Specifically, we partition the BS antenna array into~$C$ equisized clusters, each consisting of $S=B/C$ adjacent BS antennas. The corresponding partitioning of the receive vector is denoted $\bmy = \tp{[\tp{\bmy_1}, \dots, \tp{\bmy_C}]}$, where $\bmy_c\in\complexset^S, c=1,\dots,C$.
Beam-slicing then transforms each cluster into the beam-slice domain as follows:
\begin{equation}
	\hat{\bmy}_c = \bV_c\,\bmy_c, \quad c=1,\dots,C.
\end{equation}
Here, the $c$th cluster matrix $ \bV_c$ is given as a progressively phase-shifted $S$-point DFT matrix $\bF_S$ according to 
\begin{align}
	\bV_c = 
	\!\bF_S~\text{diag}\!\left(1, 
        \dots, e^{-i\frac{2\pi}{B}(c-1)(s-1)},
	\dots, e^{-i\frac{2\pi}{B}(c-1)(S-1)}
	\right).\label{eq:Vc} \end{align}
Such phase-rotated DFTs increase the ``angular diversity'' of beam-slicing to better capture the possible directions of jammers---see \fref{fig:explain_beamslicing} for a graphical explanation with two clusters. 
The action of beam-slicing is summarized as
\begin{equation}
\hat{\bmy} = \bV\bmy = \begin{bmatrix}
	\bV_1\bmy_1 \\ \vdots \\ \bV_C\bmy_C
\end{bmatrix},	
\end{equation}
where $\bV = \text{diag}(\bV_1, \dots, \bV_C)$ and $\herm{\bV}\bV=\bI_B$.
We also point out that for an (impractical) cluster size $S=B$, beam-slicing corresponds to performing a conventional beamspace transform.
In what follows, it will be convenient to 
define the \emph{beam-sliced channel matrix} $\hat\bH = \bV\bH$, and the \emph{beam-sliced jammer channel} $\hat\Hj = \bV\Hj$, which allows us to rewrite \eqref{eq:ant_io} as
\begin{align}
	\hat{\bmy} = \hat\bH \bms + \hat\Hj\sj + \hat\bmn, \label{eq:io_beamspace}
\end{align}
where $\hat\bmn=\bV\bmn$ has the same distribution as $\bmn$. 

\subsection{Data Conversion} \label{subsec:adc}
The beam-sliced signal is then converted to the digital domain.
To take into account the quantization errors of low-resolution ADCs, we assume that the beam-sliced vector $\hat{\bmy}$ is quantized as 
\begin{equation}
	\bmr = \inv{\bG}\left(\quant\left(\Re\{\bG \hat{\bmy}\}\right) + i\quant\left(\Im\{\bG \hat{\bmy}\}\right)\right). \label{eq:quantization1}
\end{equation}
Here, $\bG=\text{diag}(g_1,\dots,g_B)$ is a diagonal matrix that represents beam-wise gain-control. 
The quantization function $\quant(\cdot)$ is applied entry-wise to its input and represents a $q$-bit uniform midrise quantizer with step size $\Delta$ defined as
\begin{equation} \label{eq:quantizer}
\quant(x) \define
    \begin{cases}
      \Delta\floor{\frac{x}{\Delta}}+\frac{\Delta}{2}, & \text{if}\ |x|<\Delta 2^{q-1}\\
      \frac{\Delta}{2}(2^q-1)\frac{x}{|x|}, & \text{if}\ |x|>\Delta 2^{q-1}.
    \end{cases}
\end{equation}
For the quantizer's step size $\Delta$, we use the value which minimizes the mean-square error (MSE) between the quantizer's output~$\quant(x)$ and its input~$x$ under the assumption that $x$ is standard normal~\cite{max60a}.
For convenience, we will denote \eqref{eq:quantization1} as
\begin{equation}
	\bmr = \compquant(\hat\bmy). \label{eq:quantization2}
\end{equation}

The per-beam gains $g_b$ aim to ensure that the values entering the quantizers have unit variance per real dimension and are obtained from a set of $T$ training vectors $\tilde\bY=[\tilde\bmy_1, \dots, \tilde\bmy_T]$ as 
\begin{align}
	g_b = \sqrt{\frac{2T}{\|\tilde\bmy_{(b)}\|_2^2}},
	\quad b=1,\dots,B, \label{eq:learnG}
\end{align}
where $\tilde\bmy_{(b)}$ is the $b$th row of $\tilde\bY$.

\subsection{Jammer Interference Estimation} \label{subsec:jest}
Our soft-nulling data detection scheme (see below) treats the jamming term $\hat\Hj\sj$ in \eqref{eq:io_beamspace} as spatially correlated noise. For this, we need to know its covariance matrix $\bC_j = \Ex{\sj}{\hat\Hj \sj\herm{(\hat\Hj \sj)}}$. 
We suggest to estimate $\bC_j$ from a number of channel uses during which the UEs do not transmit any symbols, and where the jammer transmits i.i.d. jamming symbols $[\sj_1, \dots, \sj_N]$, so the quantized, beam-sliced receive matrix is
\begin{align}
	\bR_J = \compquant(\hat\bY_J) \,\,\, \text{with} \,\,\, \hat\bY_J = \hat\bmj [\sj_1, \dots, \sj_N] + \hat\bN.  \label{eq:jammer_quantiz}
\end{align}
In order to learn the jammer channel, we propose to estimate the gain matrix $\bG$ with \eqref{eq:learnG} directly from the received signals, $\bG=\bG(\hat\bY_J)$.
Our estimate $\EIf$ of the covariance matrix $\bC_j$ is
\begin{align}
	\EIf = \frac1N \bR_J\herm{\bR_J}. 
\end{align}

\subsection{Channel Estimation} \label{subsec:chest}
We estimate the UEs' channel matrix using a pilot-based least squares (LS) estimator from $U$ orthogonal pilot sequences $\bS_P = [\bms_1, \dots, \bms_U]$. 
The channel estimation pipeline passes through the beam-slicer and the quantizer. The beam-sliced receive matrix and its corresponding quantization output are
\begin{align}
	\hat\bY_P &= \hat\bH \bS_P + \hat\Hj[w_1, \dots, w_U] + \hat\bN\\
	\bR_P &= \compquant(\hat\bY_P),
\end{align}
where we estimate the gain matrix $\bG$ with \eqref{eq:learnG} from the pilot sequence itself, $\tilde\bY=\hat\bY_P$. (We fix this choice of $\bG=\bG(\hat\bY_P)$ also for the data detection phase described below.)
We then estimate the channel directly in the beam-slice domain with an LS channel estimate: 
\begin{align}
	\Hest &= \bR_P \herm{\bS_P}\inv{\left(\bS_P\herm{\bS_P}\right)} \label{eq:pilot1} \\
	&\overset{(a)}{=} \frac{1}{U\Es} \bR_P \herm{\bS_P}, \label{eq:pilot2} 
\end{align}
where $(a)$ holds because the pilot sequence is orthogonal.

\subsection{Soft-Nulling Data Detection} \label{subsec:softnulling}
Our soft-nulling data detector aims to detect the UE signal by means of linear equalization. In the detector's derivation, we will make certain idealizing assumptions under which it would be the LMMSE estimator. 
 
So far (i.e., for jammer covariance estimation and channel estimation), we have neglected the distortion introduced by the quantization step in \eqref{eq:quantization1}--\eqref{eq:quantization2}. We do not, however, neglect this distortion for the data detection step as quantization introduces distortions which are correlated with the quantizer inputs. We assume that the components of the quantizer inputs are real-valued zero-mean unit-variance Gaussian which allows us to perform a component-wise Bussgang decomposition \cite{bussgang52a} of the quantization signal as follows: 
\begin{equation}
	\quant(x) = \gamma \, x + d.
\end{equation}
Here,  $\gamma$ is the quantizer's \emph{Bussgang gain}, and the distortion $d$ has zero mean and is uncorrelated with $x$. The Bussgang gain and the second moment of the distortion are given by 
\begin{equation}
	\gamma = \frac{\Ex{}{\quant(x)x}}{\Ex{}{x^2}}	 
\end{equation}
and
\begin{equation}
	D = \Ex{}{d^2} = \Ex{}{\quant(x)^2} - \gamma^2 \Ex{}{x^2}, \label{eq:bussgang_distortion}
\end{equation}
respectively.
The Bussgang decomposition allows us to rewrite~\eqref{eq:quantization2} as
\begin{align}
	\bmr &= \compquant(\hat\bmy) \\
	&= \inv{\bG}\left(\quant\left(\Re\{\bG \hat{\bmy}\}\right) + i\quant\left(\Im\{\bG \hat{\bmy}\}\right)\right)\\
	&= \inv{\bG}\left(\gamma\,\Re\{\bG \hat{\bmy}\} + \bmd_r + i\left(\gamma\,\Im\{\bG \hat{\bmy}\} + \bmd_i\right)\right) \\
	&= \gamma\, \inv{\bG} \left(\Re\{\bG \hat{\bmy}\} + i\Im\{\bG \hat{\bmy}\}\right) +\inv{\bG}( \bmd_r + i\bmd_i) \\
	&= \gamma\, \hat{\bmy} + \inv{\bG} \bmd, \label{eq:bussgang_final}
\end{align}	
where we define $\bmd = \bmd_r + i\bmd_i$. Based on \eqref{eq:bussgang_distortion}, we make the idealized assumption that the covariance matrix of $\bmd$ is
\begin{equation}
	\bC_d = \Ex{\bmd}{\bmd\herm{\bmd}} \approx 2D\,\bI_B. \label{eq:assump_diag}
\end{equation}

We now combine \eqref{eq:bussgang_final} with \eqref{eq:io_beamspace}, which yields the input-output relation
\begin{equation}
	\bmr = \gamma\hat\bH\bms + \gamma\hat\Hj\sj + \gamma\hat\bmn + \inv{\bG} \bmd.
\end{equation}

Our linear equalizer is
\begin{align}
	\bms^\star = \bW\bmr, \label{eq:snips}
\end{align}
where the matrix $\bW$ is given as
 \begin{align}
 	\bW \!
 	&= \gamma\Es\herm{\Hest}\inv{\big( \gamma^2\Es\Hest\herm{\Hest} \!+ \gamma^2\EIf \!+ \gamma^2\No\bI_B \!+ 2D\bG^{-2}\big)} \\
 	&= \frac{1}{\gamma} \herm{\Hest}\inv{\Big( \Hest\herm{\Hest} + 
 		\frac{1}{\Es}(\EIf + \No \bI_B + 2D\gamma^{-2}\bG^{-2})\Big)}. \label{eq:WF1}
 \end{align}
Here, we use again the gain control matrix acquired during the pilot phase, $\bG=\bG(\hat\bY_P)$.
If the diagonal approximation \eqref{eq:assump_diag} and the approximations $\Hest\approx\hat\bH$ and $\EIf \approx \bC_j$ were exact, then equation \eqref{eq:snips} would implement the LMMSE estimator.  


\section{Results}
\label{sec:res}

\begin{figure*}[tbp]%
\centering 
\subfigure[BER vs.~SNR]{
\includegraphics[width=0.889\columnwidth]{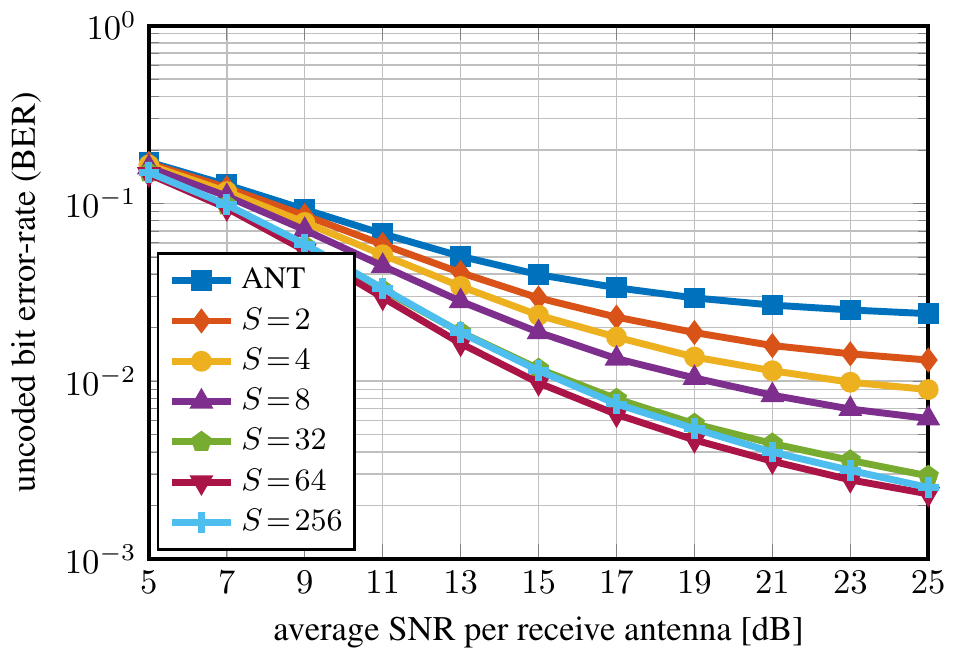}
\label{fig:sweep_S:ber}
}
~~~\quad\quad 
\subfigure[Fraction of successfully served UEs vs. SNR ]{
\includegraphics[width=0.889\columnwidth]{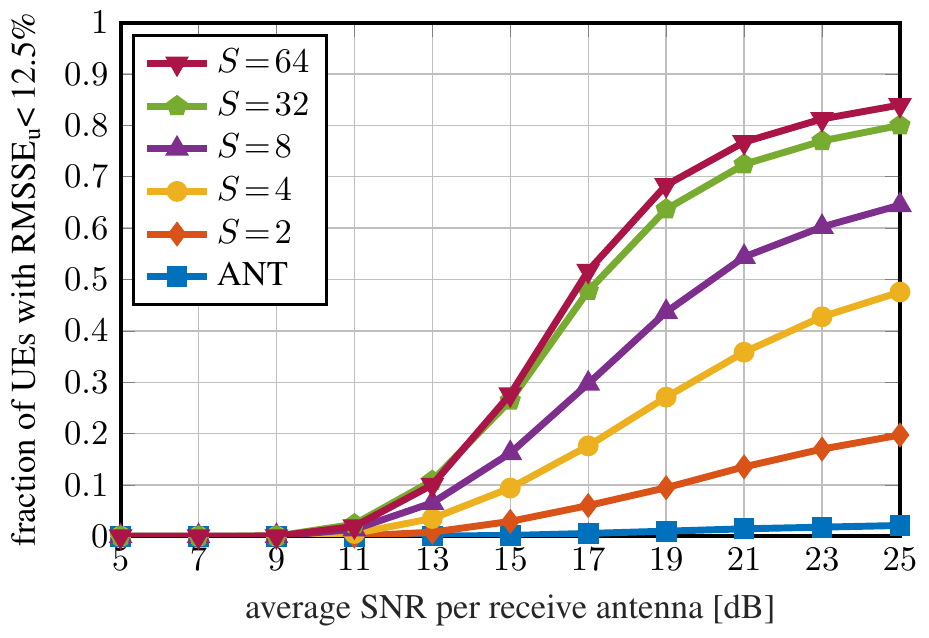}
\label{fig:sweep_S:cdf}
}
\caption{Comparison between soft-nulling in antenna domain (ANT) and SNIPS with different cluster sizes $S$, in terms of (a) uncoded bit error-rate (BER) and (b) fraction of successfully served UEs. The relative jammer power and ADC resolution are $\rho=25$\,dB and $q=4$ bits per real dimension, respectively.}
\label{fig:sweep_S}
\vspace{-0.15cm}
\end{figure*}

We now demonstrate the efficacy of SNIPS by comparing it with a baseline which differs from SNIPS only in lacking the analog beam-slicing stage. 
We note that the operation of this baseline corresponds to SNIPS with cluster size $S=1$, which implies $\bV=\bI_B$:
This baseline performs A/D-conversion and soft-nulling directly in the antenna domain.
We will show that in the presence of a strong jammer, beam-slicing with a cluster size of $S=2$ already yields significant improvements over this baseline.
We will also analyze in which cases a beam-slicer is required considering different ADC resolutions and levels of jamming power. 

\subsection{Simulation Setup and Performance Metrics}
We simulate a mmWave massive MIMO system in which $U=32$ single-antenna UEs transmit to a $B=256$ antenna BS under line-of-sight (LoS) conditions.
The channel matrices are generated using the QuaDRiGa mmMAGIC UMi model~\cite{jaeckel2014quadriga} for a 60\,GHz carrier and a uniform linear array (ULA) with $\lambda/2$ spacing. We let the $U$ UEs and the jammer be randomly placed at distances from $10$\,m to $100$\,m within a $120$\textdegree~angular sector in front of the BS. The minimum angular separation between two UEs, as well as between the jammer and any UE, is $1$\textdegree. We assume $\pm3\,$dB per-UE power control, so that the ratio between maximum and minimum per-UE-receive power is $4$. The transmit constellation is 16-QAM. 
In our simulations, we define the average receive signal-to-noise ratio (SNR) as 
\begin{align}
\textit{SNR} \define \frac{\Ex{\bms}{\|\bH\bms\|_2^2}}{\Ex{\bmn}{\|\bmn\|_2^2}}.
\end{align}
To quantify the jammer's power in comparison to a single UE, we define the relative jammer power $\rho$ as
\begin{align}
\rho \define \frac{U\Ex{\sj}{\|\Hj\sj\|_2^2}}{\Ex{\bms}{\|\bH\bms\|_2^2}} =  \frac{U\Ej \|\Hj\|_2^2}{\Es\|\bH\|_F^2}.
\end{align}

We will consider two performance metrics: Uncoded bit error-rate (BER) and a metric introduced in~\cite{Song21} called the per-UE root mean-square symbol error (RMSSE). The RMSSE for the $u$th UE over $n$ data symbol slots is defined as: 
\begin{align}
	\rmmse_u \define \sqrt{ \frac{\sum_{k=1}^{n}\big|s^\star_{u,k} - s_{u,k} \big|^2 }
	{\sum_{k=1}^{n} \left|s_{u,k} \right|^2} }. \label{eq:rmmse}
\end{align}
Here, $s_{u,k}$ and $s^\star_{u,k}$ are the transmitted and estimated data symbol of the $u$th UE at time slot $k$, respectively.
To understand the relevance of $\rmmse_u$ as a performance metric, it is helpful to compare it to the error vector magnitude (EVM) requirements in the 3GPP 5G NR technical specification \cite{3gpp21a}. The EVM is loosely speaking the square root of the sum of $\rmmse_u$-squared over all $U$ UEs. Vice versa, the $\rmmse_u$ is loosely speaking a single-UE proxy for EVM. We will therefore interpret $\rmmse_u$ as a random variable and analyze its distribution by means of Monte-Carlo simulations. 
For $16$-QAM transmission, the 3GPP 5G NR technical specification requires an EVM below $12.5$\% \cite[Tbl. 6.5.2.2-1]{3gpp21a}.
We therefore consider as a performance metric the fraction of UEs (averaged over UE placements/channel realizations, noise and jammer realizations, and data transmissions) for which the $\rmmse_u$ is below $12.5$\%.
\subsection{The Efficacy of Beam-Slicing} \label{sec:res_eff}
\fref{fig:sweep_S} evaluates the performance of SNIPS for different antenna cluster sizes $S$. 
We compare the baseline, which performs soft-nulling in antenna domain, against SNIPS with cluster sizes $S=\{2,4,8,16,32,64,256\}$, hence considering analog beam-slicing that only operates on a pair of adjacent antennas up to a single cluster consisting of the whole antenna array.
We note that with a cluster size $S=B=256$, beam-slicing corresponds to performing a beamspace transform, which is difficult to implement in the analog domain. 

The evaluation is done for a strong relative jammer power $\rho=25$\,dB and a BS with ADCs with a resolution of $q=4$~bits (per real dimension).

In \fref{fig:sweep_S:ber}, we see that beam-slicing with two-antenna clusters ($S=2$) already gives noticeable BER improvements over the antenna-domain baseline.
\fref{fig:sweep_S:ber} also shows that large clusters are superior to small ones.
However, the performance of a full beamspace transform ($S\!=\!256$) is inferior to SNIPS with $S=64$, which exhibits the best performance. The reason for this has nothing to do with how the beamspace transform distributes the jammer interference to the ADCs. Instead, the performance decrease can be explained with the fact that, after a full beamspace transform, the UE signals are concentrated to only a few ADCs, so that low-resolution ($q=4$) ADCs can no longer represent them appropriately. 
This observation suggests that the fully-centralized beamspace transform, which in any case is impractical, may not necessarily be optimal for achieving the full potential of SNIPS. For this reason, we do not consider it in the following experiments. 

The behavior of the fraction of UEs whose $\rmmse_u$ is below $12.5\%$ at a given SNR is shown in \fref{fig:sweep_S:cdf}.
In terms of this criterion, the antenna-domain baseline is unable to serve a significant fraction of UEs regardless of SNR.
In contrast,  beam-slicing with two-antenna clusters ($S=2$) serves almost $20\%$ of UEs at high SNR. 
This fraction increases with the cluster size $S$, with SNIPS being able to serve more than $80\%$ of UEs at high SNR for $S=32$ and $S=64$. 

\begin{figure}[tp]
\centering
\includegraphics[width=0.889\columnwidth]{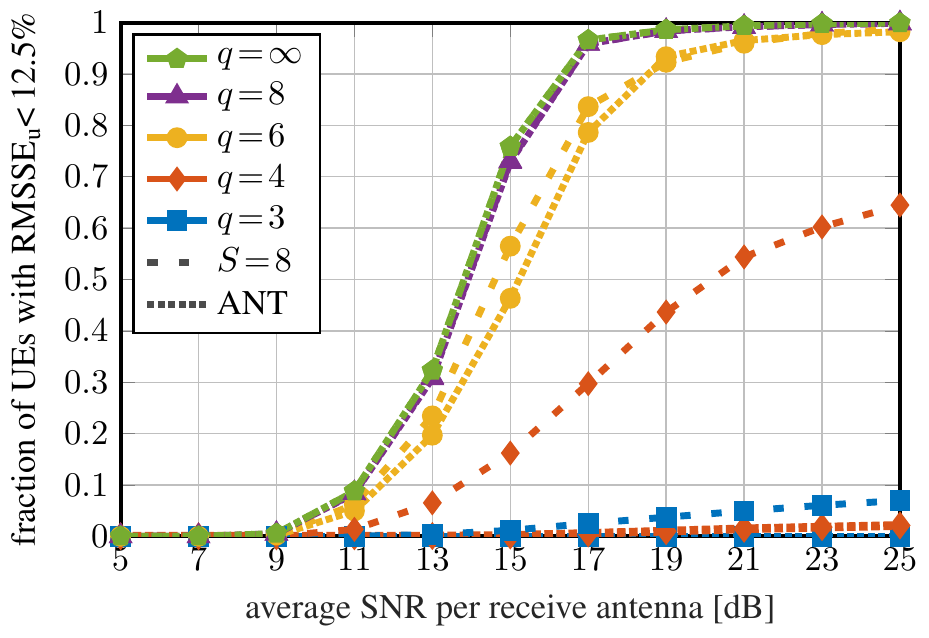}
\caption{Fraction of successfully served UEs vs.~SNR, compared between antenna domain (ANT; dotted curves) and SNIPS with $S=8$ (dashed curves) for different ADC resolutions $q$ [bits per real dimension]. The relative jammer power is $\rho=25$\,dB.}
\label{fig:sweep_ADC}
\end{figure}

\begin{figure}[tp]
\centering
\includegraphics[width=0.889\columnwidth]{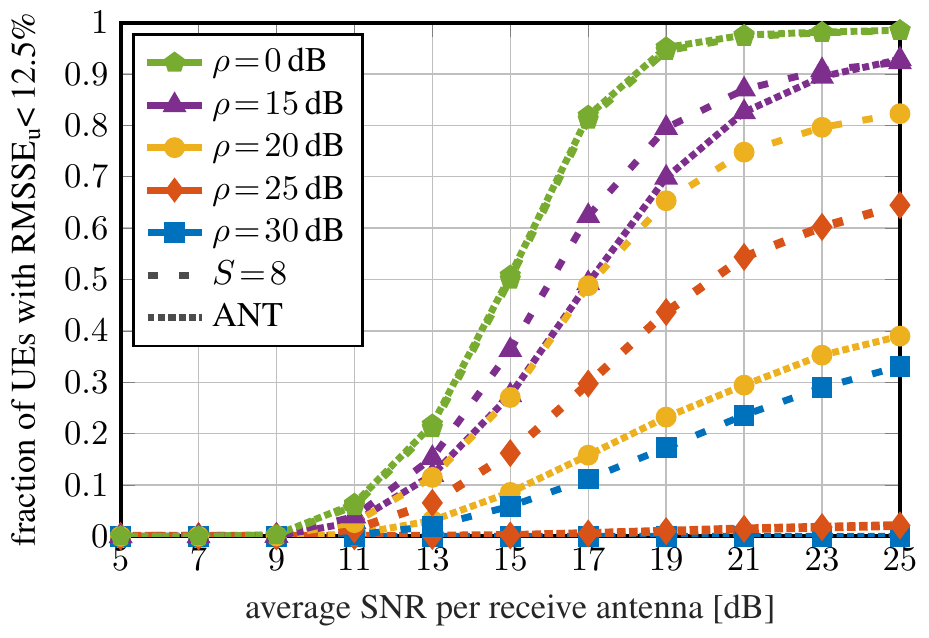}
\caption{Fraction of successfully served UEs vs.~SNR, compared between antenna domain (ANT; dotted curves) and SNIPS with $S=8$ (dashed curves) for different relative jammer powers $\rho$ [dB]. The ADC resolution is $q=4$ bits per real dimension.}
\label{fig:sweep_jammer}
\end{figure}

\subsection{When is Beam-Slicing Needed?}
The experiments in \fref{sec:res_eff} indicate that, for a strong jammer, SNIPS outperforms the antenna-domain baseline and that the best performance is achieved with a large cluster size $S$.
However, analog transforms spanning a large number of antennas are difficult to implement in practice~\cite{JoshiThesis}, with the fully-centralized beamspace transform being likely infeasible for massive MU-MIMO. 
We therefore consider a moderately-sized $S=8$ beam-slicer for our subsequent evaluations.
 
In \fref{fig:sweep_ADC}, we analyze the impact of ADC resolution for a relative jammer power $\rho=25$\,dB. We consider the fraction of UEs successfully served in terms of the criterion $\rmmse_u<\!12.5$\%.
For infinite- or high-resolution (\mbox{$q=8$}) ADCs, SNIPS and the antenna-domain baseline have identical performance. However, already for 6-bit ADCs, SNIPS outperforms the baseline. For low-resolution ADCs with $q\leq4$, the antenna-domain baseline is unable to serve a significant fraction of UEs (less than~$2$\%). In contrast, SNIPS can at least serve some UEs at high SNR even for $q=3$, and it can serve up to $65$\% of UEs for $q=4$.

In \fref{fig:sweep_jammer}, we consider the impact of the relative jammer power $\rho$ for 4-bit ADCs. We again compare SNIPS with clusters of size $S=8$ against the antenna-domain baseline. We see that their performance is virtually identical for weak jammers that are as strong as the average UE ($\rho=0$\,dB). However, for a jammer with \mbox{$\rho=15$\,dB,} SNIPS already significantly outperforms the antenna-domain baseline in terms of successfully served UEs,~and this gap continues to widen as jamming power increases further. 

These experiments confirm that strong jammers pose a serious problem for classical all-digital jamming suppression methods when combined with low-resolution ADCs. Our results also demonstrate that SNIPS is able to successfully mitigate jammer interference in a practical manner. 

\vfill



\end{document}